\documentclass[%
floatfix,
 reprint, 
superscriptaddress,
showpacs,
preprintnumbers,
nofootinbib,
 amsmath,amssymb,
 aps, prl, 
]{revtex4-2}

\usepackage[destlabel]{hyperref}

\usepackage{graphicx}
\usepackage{dcolumn}
\usepackage{bm}

\hypersetup{hidelinks}
\usepackage{booktabs}

\newcommand{\vct}[1]{\boldsymbol{#1}}
\usepackage[normalem]{ulem} 
\usepackage{cancel}
\usepackage{color}

\makeatletter
\AtBeginDocument{%
  \let\fconf@saved@maketitle\maketitle
  \let\fconf@saved@author\author
  \let\fconf@saved@affiliation\affiliation
  \let\fconf@saved@email\email
  \let\fconf@saved@collaboration\collaboration
  \let\fconf@saved@thanks\thanks
  \let\fconf@saved@abstract\abstract
  \let\fconf@saved@endabstract\endabstract
  \let\fconf@saved@hangfrom@section\@hangfrom@section
  \let\fconf@saved@sectioncntformat\@sectioncntformat
}
\newcommand\fconf@reinitfrontmatter{%
  \let\maketitle\fconf@saved@maketitle
  \let\author\fconf@saved@author
  \let\affiliation\fconf@saved@affiliation
  \let\email\fconf@saved@email
  \let\collaboration\fconf@saved@collaboration
  \let\thanks\fconf@saved@thanks
  \let\abstract\fconf@saved@abstract
  \let\endabstract\fconf@saved@endabstract
  \frontmatter@init
  \let\@hangfrom@section\fconf@saved@hangfrom@section
  \let\@sectioncntformat\fconf@saved@sectioncntformat
}
\makeatother

\begin{document}
\preprint{{RIKEN-iTHEMS-Report-26}, {FQSP-2026-3}, {YITP-26-155}}

\title{\texorpdfstring{$N\Omega$}{NOmega} Dibaryon near Unitarity from Physical-Point Lattice QCD}%

\author{Liang Zhang}
\email{zhang\_liang@fudan.edu.cn}
\affiliation{Key Laboratory of Nuclear Physics and Ion-beam Application (MOE), Institute of Modern Physics, Fudan University, Shanghai 200433, China}
\affiliation{RIKEN Center for Interdisciplinary Theoretical and Mathematical Sciences  (iTHEMS), Wako, 
351-0198, Japan}%

\author{Yan Lyu}
\email{yan.lyu@riken.jp}
\affiliation{RIKEN Center for Interdisciplinary Theoretical and Mathematical Sciences  (iTHEMS), Wako, 
351-0198, Japan}%

\author{Sinya Aoki}
\email{saoki@yukawa.kyoto-u.ac.jp}
\affiliation{Fundamental Quantum Science Program (FQSP), TRIP Headquarters, RIKEN, Wako 351-0198, Japan}
\affiliation{Center for Gravitational Physics and Quantum Information, Yukawa Institute for Theoretical Physics, Kyoto University, Kyoto 606-8502, Japan}

\author{Takumi Doi}%
\email{doi@ribf.riken.jp}
\affiliation{RIKEN Center for Interdisciplinary Theoretical and Mathematical Sciences  (iTHEMS), Wako, 
351-0198, Japan}%

\author{Tetsuo Hatsuda}
\email{thatsuda@riken.jp}
\affiliation{RIKEN Center for Interdisciplinary Theoretical and Mathematical Sciences  (iTHEMS), Wako, 351-0198, Japan}
\affiliation{Kavli Institute for the Physics and Mathematics of the Universe (Kavli IPMU), WPI,  
The University of Tokyo, Kashiwa, 
277-8568, Japan}%

\collaboration{HAL QCD Collaboration}

\author{Song Zhang}
\affiliation{Key Laboratory of Nuclear Physics and Ion-beam Application (MOE), Institute of Modern Physics, Fudan University, Shanghai 200433, China}%
\affiliation{Shanghai Research Center for Theoretical Nuclear Physics, NSFC and Fudan University, Shanghai 200438, China}

\author{Yu-Gang Ma}
\email{mayugang@fudan.edu.cn}
\affiliation{Key Laboratory of Nuclear Physics and Ion-beam Application (MOE), Institute of Modern Physics, Fudan University, Shanghai 200433, China}%
\affiliation{Shanghai Research Center for Theoretical Nuclear Physics, NSFC and Fudan University, Shanghai 200438, China}
\affiliation{School of Physics, East China Normal University, Shanghai 200241, China}


\begin{abstract}
The $N\Omega$ system is a promising dibaryon candidate because its constituents share no valence quarks and therefore avoid valence-quark Pauli repulsion. We determine the $^{5}S_{2}$ $N\Omega$ interaction in (2+1)-flavor lattice QCD at physical quark masses ($m_\pi\simeq137$ MeV) using the HAL QCD method. The interaction is attractive over the resolved range and supports a shallow state near unitarity. A direct $T$-matrix pole analysis gives 
$B_{n\Omega} = 0.61\,(^{+0.41}_{-0.30})\,(^{+0.60}_{-0.18})$~MeV; including finite-charge-size Coulomb attraction yields $B_{p\Omega} = 1.25\,(^{+0.46}_{-0.39})\,(^{+0.69}_{-0.25})$~MeV. These results provide a first-principles benchmark for future $N\Omega$ femtoscopy and dibaryon searches.
\end{abstract}

\maketitle


\noindent
{\it Introduction}:
Experimentally, the deuteron remains the only firmly established bound state of two baryons.
The $N\Omega$ system is among the most promising candidates for an additional state. 
It can form a quasi-bound state below the $N\Omega$ threshold but above
the thresholds of lower-lying octet--octet baryon channels.
Since the nucleon and the $\Omega$ baryon do not share any valence quarks
($uud$--$sss$ or $udd$--$sss$), the strong short-range repulsion associated
with Pauli exclusion at the valence-quark level is absent.
The $N\Omega$ interaction may therefore remain attractive even at short
distances, providing a favorable condition for the formation of a quasi-bound state.

The possibility of an $N\Omega$ dibaryon was already discussed in an early
quark-model study by Goldman \textit{et al.} in 1987~\cite{Goldman:1987ma}.
A first lattice QCD study of the spin-2 $N\Omega$ system was carried out
by the HAL QCD Collaboration in 2014~\cite{HALQCD:2014okw}, followed by
studies based on constituent quark models~\cite{Huang:2015yza}.
In 2016, it was pointed out that the $p\Omega$ interaction could be
tested experimentally through femtoscopic correlations in relativistic
heavy-ion collisions~\cite{Morita:2016auo}.
The $N\Omega$ interaction was subsequently investigated in chiral effective
field theory~\cite{Haidenbauer:2017sws} and in a meson-exchange model
including couplings to inelastic channels~\cite{Sekihara:2018tsb}.

In 2019, the HAL QCD Collaboration calculated
the $^5S_2$ $N\Omega$ interaction near the physical point and found a
shallow quasi-bound state~\cite{HALQCD:2018qyu}.
In the same year, the STAR Collaboration reported the first measurement of the
$p\Omega$ correlation function in Au+Au collisions at
RHIC~\cite{STAR:2018uho}.
The ALICE Collaboration subsequently measured the corresponding correlation at the LHC, with the low relative-momentum behavior found to be close to the lattice QCD prediction~\cite{ALICE:2020mfd}.
These developments stimulated further studies of $N\Omega$ femtoscopy and
the production of $\Omega$ dibaryons in relativistic heavy-ion
collisions~\cite{Morita:2019rph,Zhang:2020dma,Zhang:2021vsf,
iMendez:2024eyq,Yan:2024aap,Chen2025,Ping2}, as well as a proposal for a dedicated experimental study
at J-PARC~\cite{Aoki:2021cqa}.
Further theoretical investigations of the $N\Omega$ spectrum and
near-threshold dynamics have also appeared~\cite{Chen:2021hxs,Feng:2026wbl}.
Most recently, STAR has measured the $p\Omega$ femtoscopic correlation
in isobar collisions~\cite{Fu:2026imv}.

Because the $N\Omega$ system lies close to unitarity, its near-threshold pole is particularly sensitive to small changes in the interaction. A determination at physical quark masses is therefore important for reliably locating the near-threshold pole. Recently, the HAL QCD Collaboration generated a new set of $(2+1)$-flavor lattice QCD configurations at the physical point ($m_{\pi}\simeq137$ MeV) on the Fugaku supercomputer~\cite{Aoyama:2024cko}, enabling such a first-principles determination. On the experimental side, femtoscopy is now being used not only to constrain hadron--hadron interactions but also to infer properties of weakly bound states, as recently demonstrated by STAR for the $d\Lambda$ system~\cite{STAR:2025jwe}. 
Together with the recent experimental developments, these advances
motivate a first-principles determination of the near-threshold
$N\Omega$ pole directly at physical quark masses.

In this work, we determine the $^5S_2$ $N\Omega$ interaction directly at physical quark masses using the HAL QCD method, extract the low-energy scattering parameters, and locate the near-threshold pole on the physical sheet directly from the $T$ matrix. For $p\Omega$, we further include the Coulomb interaction with finite charge-size corrections.

\ 

\noindent
{\it HAL QCD method}:
To investigate the interaction between two baryons, we consider the four-point correlation function,
\begin{equation}
\label{eq:4pt-corr}
\begin{aligned}
    F^J(\vct{r},t)=P[\vct{r}]\sum_{\vct{x}}\left\langle 0\right|B_1(\vct{x},t)B_2(\vct{x}+\vct{r},t)\mathcal{P}^J\bar{\mathcal{J}}_{\text{wall}}(0)\left|0 \right\rangle,
\end{aligned}
\end{equation}
where $P[\vct{r}]$ and $\mathcal{P}^J$ denote a spatial projection and a total angular momentum ($J$) projection, respectively, 
and $\bar{\mathcal{J}}_{\text{wall}}(0)$ is a six-quark wall-source operator~\cite{HALQCD:2018gyl}.\footnote{Optimized source constructions~\cite{Lyu:2025lnd,Lyu:2025ncq} have recently been proposed to further enhance the signal quality.}

For the $N\Omega$ ($^5S_2$) system, the baryon operators $B_1$ and $B_2$ are taken as the nucleon and $\Omega$ interpolating fields,
\begin{equation}
\label{eq:BaryonOp}
    \begin{aligned}
        &\left[N\right]_{\alpha}(\vct{x}, t)=\epsilon_{abc}\left(u^{a^T}(\vct{x}, t){\rm C}\gamma_5d^b(\vct{x}, t)\right)q_\alpha^c(\vct{x}, t),\\
        &\left[\Omega\right]_{\beta l}(\vct{x}, t)=\epsilon_{abc}\left({s^{a^T}}(\vct{x}, t){\rm C}\gamma_ls^b(\vct{x}, t)\right)s_\beta^c(\vct{x}, t),
    \end{aligned}
\end{equation}
where $N = (p, n)^T$ and $q = (u,d)^T$. 
The total spin is projected to $s=2$ at both the sink and source, with an average taken over $s_z$ components. The spatial projection employs the $A_1^+$ irreducible representation of the cubic group, which predominantly isolates the $S$-wave component.

The $^3S_1$ $N\Omega$ channel couples strongly to $S$-wave lower-lying octet-octet channels and requires the coupled-channel formalism~\cite{Aoki:2012bb}. In contrast, the $^5S_2$ channel couples only weakly to $D$-wave octet-octet channels
~\cite{Sekihara:2018tsb,HALQCD:2018qyu}, justifying a single-channel treatment~\cite{Ishii:2006ec,Aoki:2009ji,Aoki:2020bew}.

To enhance the signal-to-noise ratio, we apply the time-dependent HAL QCD method~\cite{Ishii:2012ssm,Miyamoto:2019jjc}, which replaces the four-point correlator (Eq.~(\ref{eq:4pt-corr})) with the ``$R$-correlator''
\begin{equation}
\label{eq:R-corr}
    R^{J=2}_{N\Omega}(\vct{r},t)\equiv\frac{F^{J=2}_{N\Omega}(\vct{r},t)}{G_N(t)G_\Omega(t)},
\end{equation}
where $G_B(t)\propto\exp(-m_B t)$ is the single-baryon two-point function for a baryon of mass $m_B$.
The $R$-correlator satisfies the integro-differential equation
\begin{equation}
\label{eq:S-eq-nonlocal}
    \mathcal{K}R(\vct{r},t) = \int{U(\vct{r},\vct{r}')R(\vct{r}',t) d\vct{r}'},
\end{equation}
with $\mathcal{K}=\left(\vct{\nabla}^2/2\mu - \partial_t + (1+3\delta^2) \partial_t^2/8\mu + \mathcal{O}(\delta^2\partial_t^3/\mu^2)\right)$, where $\mu$ is the reduced mass and $\delta=(m_N-m_{\Omega})/(m_N+m_\Omega)$.
Expanding the non-local potential via the derivative expansion, $U(\vct{r},\vct{r}')=\sum_n V_n(\vct{r})\nabla^n\delta(\vct{r}-\vct{r'})$, the leading-order approximation gives
\begin{equation}
    \label{eq:Vc}
   V_{\rm LO}(\vct{r})
   =\frac{\mathcal{K}R(\vct{r},t)}{R(\vct{r},t)}.
\end{equation}
To further suppress higher orbital angular momentum contamination within the $A_1^+$ projection, we apply an additional $S$-wave projection via the Misner method~\cite{Misner:1999ab,Miyamoto:2019jjc}, after which the $R$-correlator reduces to $R(r,t)$ and the Laplacian can be evaluated analytically rather than by finite difference on the lattice. 

\

\noindent
{\it Lattice setup}: 
The gauge configurations, collectively called the ``HAL-conf-2023'' ensemble (``F-conf'' for short), 
were generated on the Fugaku supercomputer at RIKEN 
using $(2+1)$-flavor lattice QCD with the Iwasaki gauge action at $\beta=1.82$ and nonperturbatively $O(a)$-improved Wilson quarks with stout smearing~\cite{Aoyama:2024cko}.
The lattice spacing, $a\simeq 0.084372$~fm ($a^{-1}\simeq 2338.8$~MeV), was determined using the experimental
$\Omega$ baryon mass, $m_{\Omega}=1672.45$~MeV, 
as input~\cite{Aoyama:2024cko, ParticleDataGroup:2024cfk}.
The $96^4$ lattice thus has a spatial extent $L\simeq 8.1$~fm. 
Importantly, F-conf realizes the physical point with $m_{\pi}\simeq137.1$~MeV and $m_K\simeq501.8$~MeV, allowing the $N\Omega$ interaction to be determined directly at physical quark masses without a quark-mass extrapolation.
The nucleon mass, 
$m_N=939.7(1.8)(^{+0.2}_{-1.7})$~MeV,
taken from the F-conf study~\cite{Aoyama:2024cko},
is also consistent with the PDG
value~\cite{ParticleDataGroup:2024cfk}
and is used in the analysis below.

The analysis employs 1600 configurations, each with 384 measurements including forward/backward propagation averaging, four lattice rotations exploiting hypercubic symmetry, and 48 distinct source locations. 
The configurations are grouped into 20 bins of 80 each to mitigate autocorrelation effects.
We have checked that the results are almost unchanged when using 10 or 40 bins.
Jackknife estimates are employed for the potential $V_{\rm LO}(\vct{r})$, the scattering parameters $a_0$ and $r_{\text{eff}}$, and the binding energies. 


\

\noindent
\textit{Results for the potential}:
We extract the potential via Eq.~\eqref{eq:Vc}, truncating the
time-derivative terms at $\mathcal{O}({\partial_t^2/\mu})$, 
where the ${\cal O}(\delta^2 \partial_t^3/\mu^2)$ term is found to be negligible.
At small $t$,
contamination from inelastic excited states may distort the signal, while
at large $t$, statistical fluctuations dominate. We therefore use the window
$12 \leq t/a \leq 16$ and take the spread of the potential over this window
as the systematic uncertainty associated with the residual $t$ dependence.

\begin{figure}[t]
    \includegraphics[width=1\linewidth]{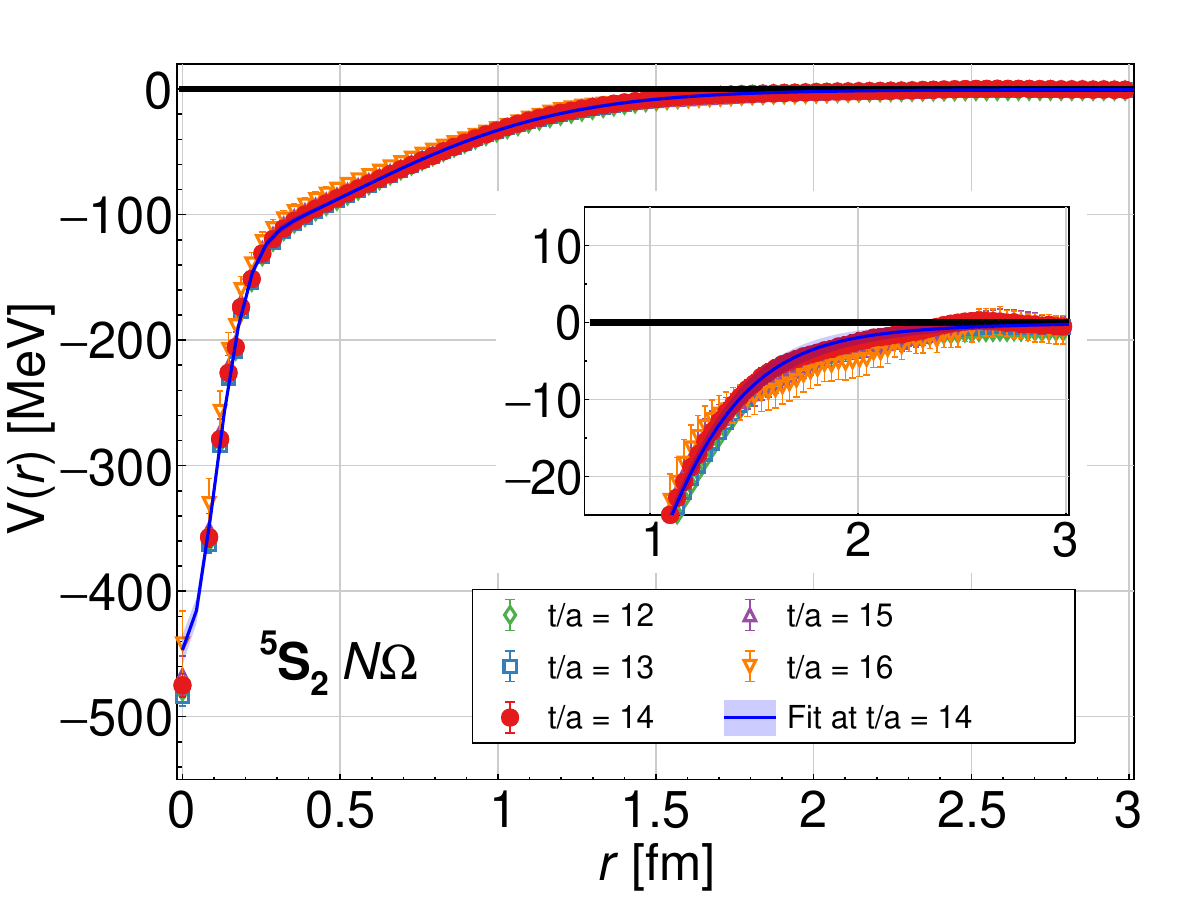}
    \caption{The $^5S_2$ $N\Omega$ central potential  extracted at five time slices,
    $t/a=12$--$16$.
    The potential is attractive over the statistically resolved range and exhibits a pronounced attractive core ($r<0.3$ fm), a mid-range shoulder ($0.3<r<1.5$ fm) and a long-range attractive tail ($r>1.5$ fm). The blue curve is a fit to the $t/a = 14$ data over the range $0 < r < 3$~fm, using Eq.~(\ref{eq:Gaus-Guas-TPE}). The blue band indicates the statistical uncertainty of the fit.}
    \label{fig:potential}
\end{figure}

Figure~\ref{fig:potential} shows the $^5S_2$ $N\Omega$ potential.
The potential is attractive over the statistically resolved range, a common feature in hadron-pair systems that do not share valence quarks~\cite{HALQCD:2014okw,HALQCD:2018qyu,Lyu:2022imf,Lyu:2024ttm,Zhang:2025zaa}. The absence of valence-quark Pauli blocking removes an important mechanism for short-range repulsion. The spin-2 $N\Omega$ potential is consistent across all five time slices within statistical errors, indicating that this channel couples only weakly to other baryon-pair channels below the $N\Omega$ threshold~\cite{Sekihara:2018tsb,HALQCD:2018qyu,Feng:2026wbl}. In contrast, we find that the spin-1 channel exhibits pronounced time dependence, signaling strong coupling to lower-lying baryon-pair channels and necessitating a coupled-channel HAL QCD analysis.

The high precision resolves three characteristic regions of the
interaction: an attractive short-range core ($r<0.3$ fm), an
intermediate-range shoulder ($0.3<r<1.5$ fm), and a long-range
attractive tail beyond $1.5$ fm.  
Guided by this structure, we fit the potential with a sum of two Gaussians to capture the core and shoulder, plus a 
two-pion-exchange (TPE) tail for the long-range 
behavior~\cite{HALQCD:2018qyu},
\begin{equation}
    \label{eq:Gaus-Guas-TPE}
    V_{\text{fit}}(r)=\sum_{i=1}^2 a_i e^{-b_i r^2} + a_3 \left(1-e^{-(r/R)^2}\right)\left(\frac{e^{-m_{\pi}r}}{r}\right)^2.
\end{equation}
The form factor $(1-e^{-(r/R)^2})$ in Eq.~\eqref{eq:Gaus-Guas-TPE} suppresses the 
TPE contribution at short range, where the Gaussian terms dominate. Guided by the nucleon radius ($\sim 1$~fm), we set $R=2$~fm. The pion mass is fixed to the lattice value $m_\pi = 137.1$~MeV, and the remaining parameters are fitted to the data at $r < 3$~fm.
The fit at $t/a=14$ is shown as a representative example in Fig.~\ref{fig:potential}, and the parameters for all time slices are collected in Table~\ref{tab:fit-params}.

We also study
the long-range behavior of the potential
by extracting the spatial effective energy,
$E_{\rm eff}(r) = -\{\ln[V_{\rm LO}(r)r^2/a_3]/r\}$~\cite{Lyu:2022imf,Lyu:2024ttm,Lyu:2023xro}.
For $1.5\lesssim r\lesssim2.5$~fm, $E_{\rm eff}(r)$ is compatible, within its large statistical uncertainty ($\sim100$~MeV), with $2m_\pi$, as expected for a TPE tail~\cite{HALQCD:2018qyu,Lyu:2022imf,Lyu:2024ttm,Lyu:2023xro}. The present data, however, do not establish a statistically significant plateau, and a more quantitative analysis is left for future work.

\begin{table}[b]
\caption{Fit parameters of the $^5S_2$ $N\Omega$ potential (Fig.~\ref{fig:potential}) using Eq.~\eqref{eq:Gaus-Guas-TPE} at each time slice. The statistical errors (in parentheses) are estimated by the jackknife method.}
\label{tab:fit-params}
\footnotesize
\begin{tabular}{cccccc}
\toprule
$t/a$ & $a_1$~[MeV] & $b_1$~[fm$^{-2}$] & $a_2$~[MeV] & $b_2$~[fm$^{-2}$] & $a_3$~[MeV$\cdot$fm$^{2}$]
\\
\midrule
$12$ & $-312.7(2.2)$ & $51.8(1.1)$ & $-76.3(7.5)$ & $1.44(4)$ & $-289.0(46.9)$ \\
$13$ & $-314.8(3.9)$ & $50.6(2.3)$ & $-83.9(11.3)$ & $1.43(8)$ & $-244.1(78.1)$ \\
$14$ & $-311.5(6.7)$ & $49.0(2.4)$ & $-94.7(16.6)$ & $1.38(10)$ & $-162.6(111.1)$ \\
$15$ & $-307.1(15.6)$ & $52.6(4.4)$ & $-76.2(17.2)$ & $1.47(19)$ & $-265.3(118.3)$ \\
$16$ & $-285.4(24.1)$ & $56.4(8.9)$ & $-56.3(17.3)$ & $1.73(46)$ & $-358.8(131.5)$ \\
\bottomrule
\end{tabular}
\end{table} 

\

\begin{figure}
    \includegraphics[width=1\linewidth]{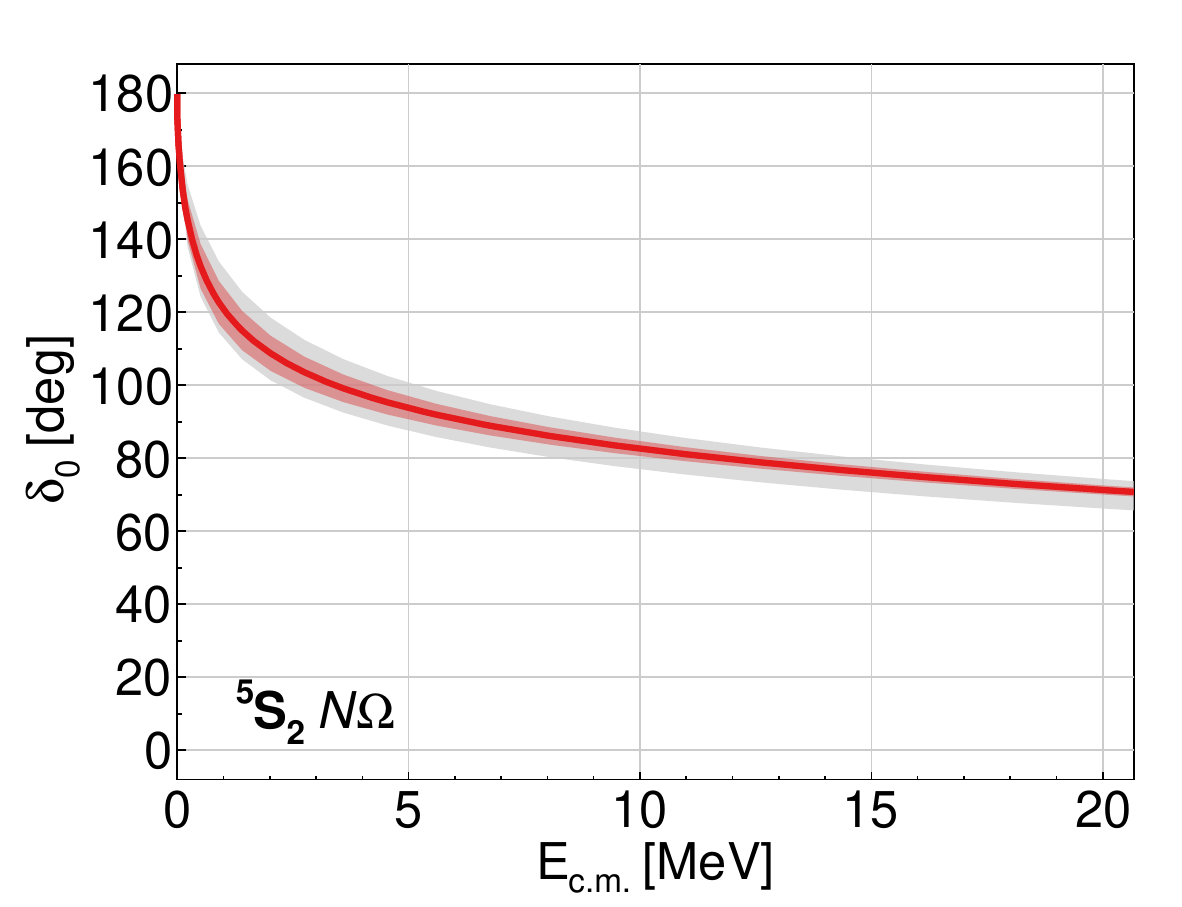}
  \caption{The $^5S_2$ $N\Omega$ scattering phase shift as a function of the center-of-mass kinetic energy $E_{\text{c.m.}}$. The red band shows the statistical error, and the gray band shows the total uncertainty, with statistical and systematic errors added in quadrature.}
  \label{fig:phaseshift}
\end{figure}

\noindent
{\it Observables}:
The $S$-wave scattering phase shift $\delta_0$ is obtained by solving the Schr\"odinger equation in infinite volume
with the fitted potential $V_{\text{fit}}$. 
The $N\Omega$ phase shift without Coulomb attraction is shown in Fig.~\ref{fig:phaseshift} as a function of the center-of-mass kinetic energy $E_{\text{c.m.}}$. 
Results from $12\leq t/a\leq16$ are found to agree within statistical errors.
We take central values and statistical uncertainties from the results at $t/a=14$, and estimate the systematic uncertainties from the spread among the five time slices. Unless otherwise noted, all numerical results below follow this procedure. 

For $p\Omega$, we account for the additional Coulomb attraction between the finite charge distributions of the proton and the $\Omega$ baryon. We write the Coulomb potential as
\begin{equation}
    \label{eq:coulomb-FC}
    V_{\rm Coul}(r)=-\frac{\alpha}{r}\,F_{\rm C}(r),
\end{equation}
where the finite-size correction $F_{\rm C}(r)$ is obtained from a dipole form factor~\cite{Can:2015exa,Lyu:2021qsh}, and its explicit form is given in Sec.~\ref{supp:coulomb-form} of the Supplemental Material~\cite{SupplementalMaterial}. We use the experimentally measured proton charge radius of $0.841$~fm~\cite{Mohr:2024kco}, consistent with recent hydrogen spectroscopy determinations~\cite{Maisenbacher:2026nau,Bullis:2026ebw}, and the lattice result of $0.571(18)$~fm for the $\Omega$ charge radius~\cite{Can:2015exa}.

As $k\to 0$, the $S$-wave phase shift for $n\Omega$ approaches $180^\circ$. Within the present single-channel description, the observed $\delta_0\to180^\circ$ as $k\to0$ implies one (quasi-)bound state pole through Levinson's theorem.
This conclusion can be verified quantitatively through the effective range expansion (ERE). A bound state corresponds to a pole of the $S$-matrix on the imaginary momentum axis, 
with the pole momentum $k=i\kappa$ ($\kappa>0$) satisfying
$k\cot\delta_0(k)-ik=0$. Substituting the ERE
\begin{equation}
    \label{eq:ERE}
    k \cot\delta_0=\frac{1}{a_0}+\frac{1}{2}r_{\text{eff}}k^2+\mathcal{O}(k^4),
\end{equation}
gives, within the truncated ERE, the conditions $a_0<0$ and $1+2r_{\text{eff}}/a_0>0$ for a shallow bound-state pole. Fitting the $n\Omega$ phase shifts to Eq.~\eqref{eq:ERE} gives
\begin{equation}
    \label{eq:nOmg-a0-reff}
    \begin{aligned}
        1/a_0&=-0.12(3)(^{+0.02}_{-0.04})~\text{fm}^{-1},\\
        r_{\text{eff}}&=1.26(2)(^{+0.14}_{-0.01})~\text{fm},
    \end{aligned}
\end{equation}
where the first set of parentheses gives the jackknife statistical uncertainty, and the second gives the systematic uncertainty.
The magnitude of the scattering length, $|a_0|\simeq 8.3$~fm, is much larger than the effective range, $r_{\text{eff}}\simeq1.26$~fm, illustrating the proximity of this system to the unitary limit.

\begin{figure}
    \includegraphics[width=1\linewidth]{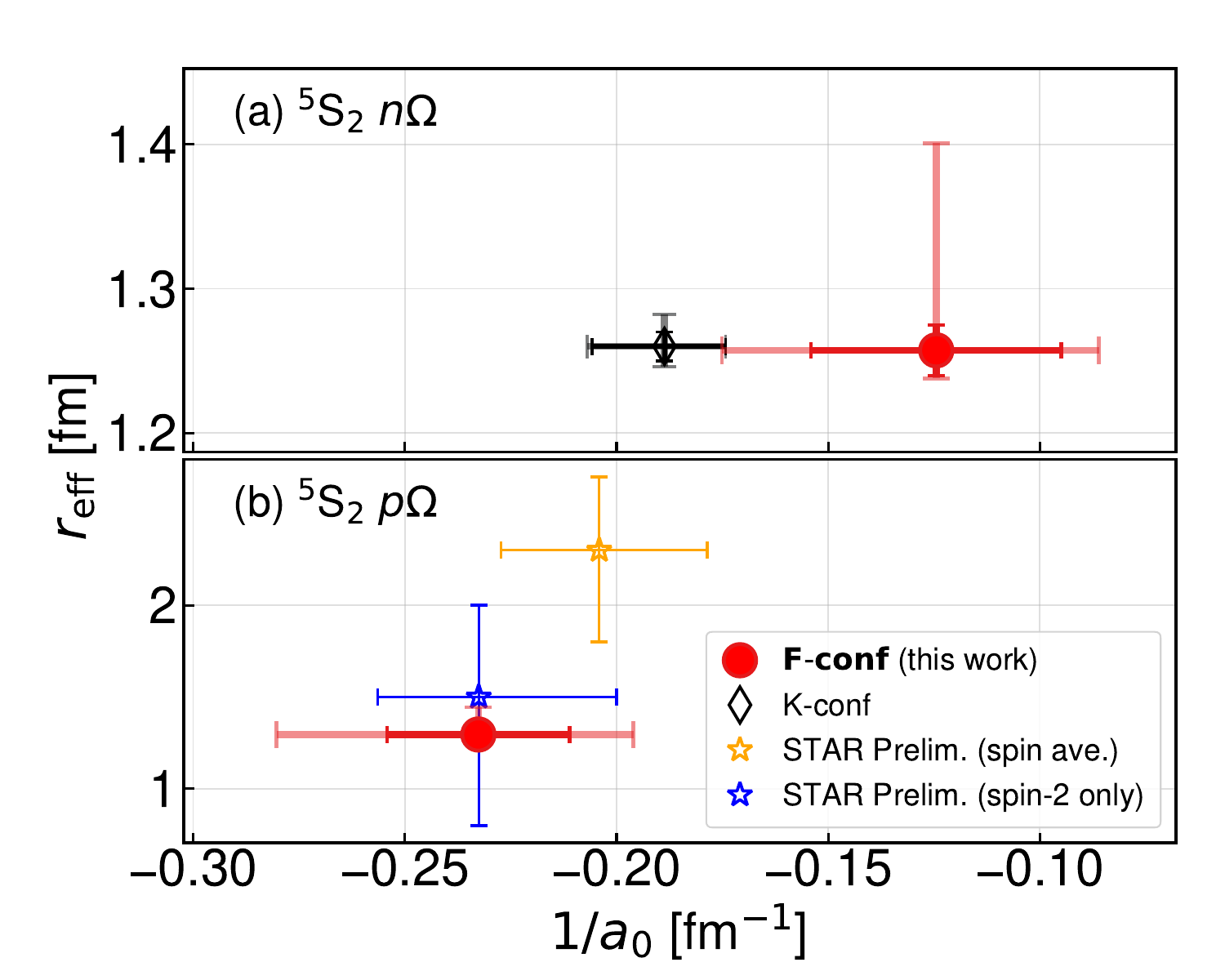}
  \caption{Scattering parameters, $1/a_0$ and $r_{\rm eff}$, of the $^5S_2$ $N\Omega$ system for (a) $n\Omega$ and (b) $p\Omega$, with the Coulomb interaction included in (b).
  The red circles and the black diamond denote the results at
  $m_{\pi}\simeq 137$~MeV (F-conf) and $m_{\pi}\simeq 146$~MeV (K-conf)~\cite{HALQCD:2018qyu}, respectively. Note that 
  the scattering parameters for the $p\Omega$ system were not calculated for the K-conf.
  The thick, dark-colored error bars indicate statistical uncertainties, whereas the thin, light-colored error bars indicate total uncertainties, with statistical and systematic errors added in quadrature.
  In (b), the statistical error of $r_{\rm eff}$ is smaller than the symbol size. The yellow 
  and blue stars show the STAR preliminary results~\cite{Fu:2026imv} obtained from the spin-averaged fit and the fit with the spin-2 channel only, respectively.}
    \label{fig:comp-lattice-exp}
\end{figure}

The values in Eq.~(\ref{eq:nOmg-a0-reff}) satisfy the bound-state criterion, which confirms
a bound-state pole within the present single-channel description. 
Note that the left-hand cut associated with TPE
is well separated from this bound-state pole. 
The scattering parameters are consistent within uncertainties with the previous HAL QCD result~\cite{HALQCD:2018qyu},
obtained using the gauge configurations
generated on the K computer 
near the physical point, $m_\pi \simeq 146$ MeV (``K-conf'')~\cite{Ishikawa:2015rho},
once the opposite sign convention for $1/a_0$ used there is taken into account (Fig.~\ref{fig:comp-lattice-exp}).

For $p\Omega$, the situation is more involved. The Coulomb attraction between oppositely charged particles guarantees a series of shallow bound states independently of the strong interaction, and Levinson's theorem does not directly apply in the presence of Coulomb attraction. To extract the strong interaction scattering parameters, we employ the Coulomb-corrected ERE~\cite{Bethe:1949yr,Kong:1999sf},
\begin{equation}
    \label{eq:coulomb-ERE}
A_{\mathrm{c}}(\eta)\,k
\left(\cot\delta_0^{\mathrm{C}}-i\right)
+2k\eta H(\eta)
=
\frac{1}{a_0^{\mathrm{C}}}
+\frac{1}{2}r_{\mathrm{eff}}^{\mathrm{C}}k^2
+\mathcal{O}(k^4).
\end{equation}
Here, $\delta^\mathrm{C}_0$ is the phase shift relative to pure Coulomb scattering, $\eta = \mu Z_1Z_2 \alpha/k$ is the Sommerfeld parameter, $Z_1$ and $Z_2$ are the particle charges, and $\alpha$ is the fine-structure constant;
$A_c(\eta)=2\pi\eta/(e^{2\pi\eta}-1)$ is the Coulomb penetration factor, and $H(\eta)=\psi(i\eta)+1/(2i\eta)-\ln(i\,\operatorname{sgn}(Z_1 Z_2)\,\eta)$ involves the digamma function $\psi$.
A fit to the $p\Omega$ phase shifts yields
\begin{equation}
    \label{eq:pOmg-a0-reff}
    \begin{aligned}
        1/a^{{\rm C}}_0 &= -0.23(2)(^{+0.03}_{-0.04})~\text{fm}^{-1},\\
        r^{{\rm C}}_{\text{eff}} &= 1.30(2)(^{+0.15}_{-0.01})~\text{fm}.
    \end{aligned}
\end{equation}
Compared to the $n\Omega$ result, the Coulomb correction reduces the magnitude of the scattering length,
while the effective range remains essentially unchanged~\cite{Jackson:1950zz,Kong:1999sf,Lyu:2021qsh}.

To determine the near-threshold pole without relying solely on the ERE, we use the Lippmann--Schwinger equation. For $n\Omega$, 
a bound state corresponds to a pole of $T=V+VG_0T$ at $E=-\kappa^2/2\mu$ ($\kappa>0$),
or to a zero of
\begin{equation}
   \det\!\left[1-VG_0(-\kappa^2/2\mu)\right]=0,
   \label{eq:fredholm}
\end{equation}
where $G_0(E)=(E-H_0)^{-1}$ is the free propagator.
For $p\Omega$, the long-range Coulomb interaction cannot be treated perturbatively.
We therefore separate the potential into the long-range Coulomb part $-\alpha/r$
and the short-range part $V_s=V+V_{\rm Coul}+\alpha/r$, where $V$ is the
lattice QCD potential in Eq.~\eqref{eq:Gaus-Guas-TPE} and $V_{\rm Coul}$
is the Coulomb potential with the finite-size correction
[Eq.~\eqref{eq:coulomb-FC}]. The corresponding Coulomb Green's function is
$G_{\rm C}(E)=(E-H_0+\alpha/r)^{-1}$. The bound-state pole is then obtained
from Eq.~\eqref{eq:fredholm} by replacing $V$ with $V_s$ and $G_0$ with
$G_{\rm C}$, with the determinant evaluated in the Coulomb scattering basis.

\begin{figure}[t]
    \includegraphics[width=1\linewidth]{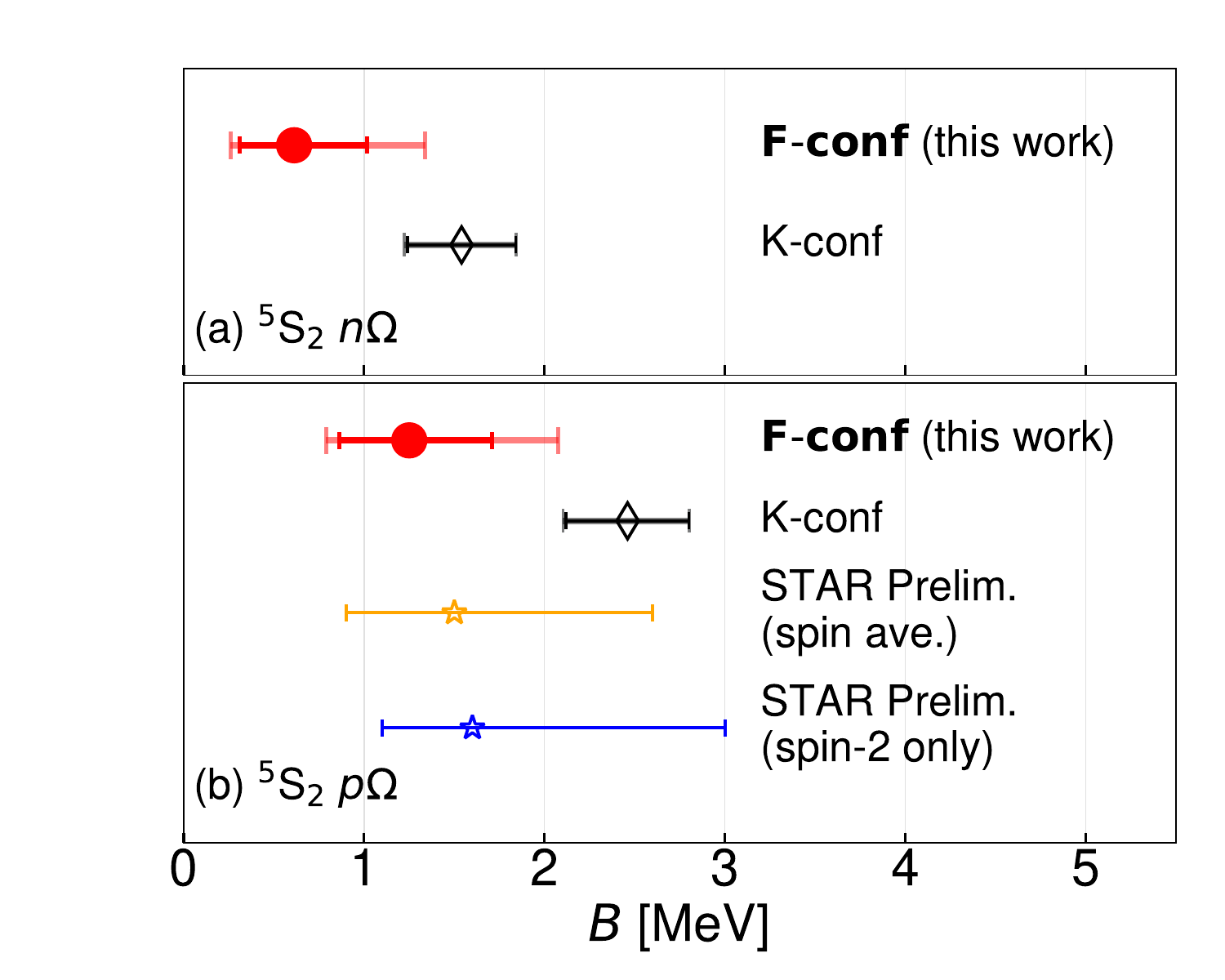}
    \caption{Comparison of the $^5S_2$ $N\Omega$ binding energies 
    for $m_{\pi}\simeq 137$~MeV (F-conf) and $m_{\pi}\simeq 146$~MeV (K-conf)~\cite{HALQCD:2018qyu}.
    The $p\Omega$ binding energy for the K-conf was obtained by solving the Schr\"odinger equation with the QCD potential plus the Coulomb potential for point-like particles.
    The STAR preliminary data without Coulomb modification are also shown~\cite{Fu:2026imv}.
    }
    \label{fig:comp-binding}
\end{figure}

The resulting pole momenta are
\begin{equation}
    \label{eq:pole-pos}
    \begin{aligned}
        \kappa_{n\Omega} &= 27.1(7.9)\,(^{+11.1}_{-4.3})~\text{MeV},\\
        \kappa_{p\Omega} &= 38.8(6.5)\,(^{+9.5}_{-4.1})~\text{MeV},
    \end{aligned}
\end{equation}
where the first uncertainty is statistical and was obtained by jackknife resampling and the second is systematic. 
For $n\Omega$, the $1\sigma$ interval of $\kappa$ lies entirely in the positive region, indicating the pole is located on the physical sheet. A bootstrap analysis of the pole momentum, which quantifies the bound-state existence for all time slices, is presented in Sec.~\ref{supp:bootstrap} of the Supplemental Material~\cite{SupplementalMaterial}.
The corresponding binding energies are
\begin{equation}
    \label{eq:binding-eng}
    \begin{aligned}
        B_{n\Omega} &= 0.61\,(^{+0.41}_{-0.30})\,(^{+0.60}_{-0.18})~\text{MeV},\\
        B_{p\Omega} &= 1.25\,(^{+0.46}_{-0.39})\,(^{+0.69}_{-0.25})~\text{MeV}.
    \end{aligned}
\end{equation}
The physical-point binding energies have smaller central values than the previous near-physical-point results~\cite{HALQCD:2018qyu} for both channels, but remain consistent within the combined statistical and systematic uncertainties (Fig.~\ref{fig:comp-binding}).

In Fig.~\ref{fig:comp-lattice-exp} and Fig.~\ref{fig:comp-binding}, the STAR preliminary results~\cite{Fu:2026imv} obtained from the spin-averaged fit and the fit with the spin-2 channel only are shown for qualitative comparison. A quantitative comparison should await the final experimental analysis, in particular because the current extraction of the binding energy does not include the Coulomb modification~\cite{Fu:2026imv}.

We tested fit stability and checked the leading-order HAL QCD
approximation by comparing finite-volume energy levels extracted from
projected correlators with predictions of the potential and by applying
L\"uscher's relation~\cite{Luscher:1990ux} (see Sec.~\ref{app:FV-energy}
of the Supplemental Material~\cite{SupplementalMaterial} and the End Matter). 
Sensitivity of the results to the short-range part of the potential is also checked to estimate the discretization errors.
These checks support
the low-energy results in this paper, while a continuum extrapolation remains necessary
to fully quantify discretization errors.


\

\noindent
{\it Summary}:
In this Letter, we have determined the $^5S_2$ $N\Omega$ interaction directly at physical quark masses in $(2+1)$-flavor lattice QCD using the HAL QCD method. The interaction is attractive over the statistically resolved range and supports a shallow bound-state pole in the single-channel approximation, close to unitarity. By directly locating the near-threshold pole of the $T$ matrix on the physical sheet, we obtain
$B_{n\Omega} = 0.61\,(^{+0.41}_{-0.30})\,(^{+0.60}_{-0.18})$~MeV and
$B_{p\Omega} = 1.25\,(^{+0.46}_{-0.39})\,(^{+0.69}_{-0.25})$~MeV, where the finite-size Coulomb attraction is included for $p\Omega$. The shallow binding and small inverse scattering length place the physical $N\Omega$ system close to the unitary limit. This physical-point result provides a first-principles benchmark for ongoing and future high-statistics $p\Omega$ femtoscopic analyses in $pp$ and nucleus--nucleus collisions, as well as for dedicated $N\Omega$ studies at J-PARC~\cite{Aoki:2021cqa}. On the theory side, a coupled-channel treatment including the lower-lying $D$-wave octet--octet channels will enable a first-principles determination of the width of the state, while the inclusion of QED and QCD isospin-breaking effects will further clarify the $n\Omega$--$p\Omega$ splitting.

\par\vspace{\baselineskip}
\begin{acknowledgments}
\noindent{\it Acknowledgments}:\enspace
We thank members of the HAL QCD Collaboration for stimulating discussions. We also thank Kehao Zhang for his helpful discussions on experimental progress in $p\Omega$ femtoscopy.
This work was supported in part by 
the JSPS KAKENHI 
(Grant Nos.~JP26K21721, JP23H05439, and JP22H00129),
%
the Japan Science and Technology Agency (JST) as part of Adopting Sustainable Partnerships for Innovative Research Ecosystem 
(ASPIRE Grant No. JPMJAP2318),
RIKEN TRIP initiative,
and
RIKEN Incentive Research Project.
The lattice QCD calculations were carried out using the Fugaku supercomputer at RIKEN through the HPCI System Research Project (hp240157, hp240213, hp250195, hp250224),  MEXT Program for Promoting Researches on the Supercomputer Fugaku 
``Simulation for basic science: from fundamental laws of particles to creation of nuclei" (JPMXP1020200105) and 
``Simulation for basic science: approaching the new quantum era" (JPMXP1020230411),
and
MEXT Next-Generation Computational Science Global Reach \& Innovation Program ``Building a Lattice QCD Ecosystem for the GPU and AI Era" (JPMXP1920260208).
We also acknowledge support from
ILDG/JLDG~\cite{ldg, Amagasa:2015zwb}
and Joint Institute for Computational Fundamental Science (JICFuS).
YGM was supported by the National Natural Science Foundation of China under contract Nos. 12688101 and 12547102, and the STCSM under Grant Nos.~23590780100 and 23JC1400200. 
\end{acknowledgments}

\bibliography{05}


\appendix
\section{End Matter}

\begingroup
\noindent\textit{Fit-function and fit-range dependence}:\enspace
To assess the dependence on the potential fit function, we vary
the parameter $R$ in Eq.~\eqref{eq:Gaus-Guas-TPE}, which controls the
short-distance part of the two-pion-exchange contribution.
Repeating the analysis with $R=1.5$, $2.0$, and $2.5$~fm, we find that
the extracted scattering parameters remain stable, with variations
negligible compared with their statistical uncertainties.
We also examine the dependence on the fit range $0<r<r_{\rm max}$ by
varying $r_{\rm max}=2.0$, $3.0$, and $4.0$~fm, and find negligible
changes in the resulting observables.
As a further check, we repeat the analysis using a purely
phenomenological fit function consisting of three Gaussians,
$V_{\text{fit}}(r)=\sum_{i=1}^3 \tilde{a}_i e^{-\tilde{b}_i r^2}$,
 instead of the fit form in  Eq.~\eqref{eq:Gaus-Guas-TPE}.
The resulting observables again remain essentially unchanged, with
variations much smaller than their statistical uncertainties.

\noindent\textit{Leading-order approximation}:\enspace
We examine possible truncation errors in the derivative expansion,
since the potential is obtained at leading order (LO),
Eq.~\eqref{eq:Vc}.
While the residual Euclidean-time dependence has been included in the
quoted systematic uncertainties, we perform an additional consistency
check using finite-volume energy levels extracted from temporal
correlators projected onto eigenfunctions of the LO Hamiltonian
~\cite{Iritani:2018vfn,Lyu:2022tsd,Lyu:2025lnd,Lyu:2025ncq}.
As detailed in Sec.~\ref{app:FV-energy} of the Supplemental Material,
these levels agree with the eigenvalues predicted by the LO potential.
Moreover, the values of $k\cot\delta_0$ inferred from these energies using
L\"uscher's relation~\cite{Luscher:1990ux} agree with those obtained
from the LO potential in infinite volume. These comparisons support the
reliability of the LO approximation for low-energy observables.

\noindent\textit{Discretization effects}:\enspace
Since the calculation is performed at a single lattice spacing,
discretization effects remain a major source of systematic uncertainty.
Because such effects are expected to be most apparent in the short-range
part of the potential, we test the sensitivity of the results to
modifications at the lattice scale, $r\lesssim a$, by replacing
$V_{\text{fit}}(r)$ with
$(1+\epsilon\exp[-r^2/a^2])V_{\text{fit}}(r)$,
where $\epsilon=\pm1$.
We find that the scattering length, effective range, and bound-state
pole momentum change by only $\sim5\%$, $\sim1\%$, and $\sim5\%$,
respectively, relative to the original results ($\epsilon=0$), with
variations smaller than their statistical uncertainties.
This test does not provide a quantitative estimate of the full
discretization error, but shows that our results are insensitive to
substantial modifications of the potential at the lattice scale.
Calculations at multiple lattice spacings will be necessary for an
explicit continuum extrapolation.
\endgroup

\clearpage
\makeatletter\fconf@reinitfrontmatter\makeatother
\setcounter{secnumdepth}{1}
\renewcommand{\theequation}{S\arabic{equation}}
\setcounter{equation}{0}
\renewcommand{\thefigure}{S\arabic{figure}}
\setcounter{figure}{0}
\renewcommand{\thetable}{S\arabic{table}}
\setcounter{table}{0}
\renewcommand{\thesection}{S\arabic{section}}
\setcounter{section}{0}
\renewcommand{\theHequation}{SM-eq\arabic{equation}}
\renewcommand{\theHfigure}{SM-fig\arabic{figure}}
\renewcommand{\theHtable}{SM-tab\arabic{table}}
\renewcommand{\theHsection}{SM-sec\arabic{section}}

\preprint{{RIKEN-iTHEMS-Report-26}, {FQSP-2026-3}, {YITP-26-155}}

\title{Supplemental Material \\
\vspace{0.5cm}
\texorpdfstring{$N\Omega$}{NOmega} Dibaryon near Unitarity from Physical-Point Lattice QCD}

\author{Liang Zhang}
\affiliation{Key Laboratory of Nuclear Physics and Ion-beam Application (MOE), Institute of Modern Physics, Fudan University, Shanghai 200433, China}
\affiliation{RIKEN Center for Interdisciplinary Theoretical and Mathematical Sciences  (iTHEMS), Wako, 
351-0198, Japan}%

\author{Yan Lyu}
\affiliation{RIKEN Center for Interdisciplinary Theoretical and Mathematical Sciences  (iTHEMS), Wako, 
351-0198, Japan}%

\author{Sinya Aoki}
\affiliation{Fundamental Quantum Science Program (FQSP), TRIP Headquarters, RIKEN, Wako 351-0198, Japan}
\affiliation{Center for Gravitational Physics and Quantum Information, Yukawa Institute for Theoretical Physics, Kyoto University, Kyoto 606-8502, Japan}

\author{Takumi Doi}%
\affiliation{RIKEN Center for Interdisciplinary Theoretical and Mathematical Sciences  (iTHEMS), Wako, 
351-0198, Japan}%

\author{Tetsuo Hatsuda}
\affiliation{RIKEN Center for Interdisciplinary Theoretical and Mathematical Sciences  (iTHEMS), Wako, 351-0198, Japan}
\affiliation{Kavli Institute for the Physics and Mathematics of the Universe (Kavli IPMU), WPI,  
The University of Tokyo, Kashiwa, 
277-8568, Japan}%

\collaboration{HAL QCD Collaboration}

\author{Song Zhang}
\affiliation{Key Laboratory of Nuclear Physics and Ion-beam Application (MOE), Institute of Modern Physics, Fudan University, Shanghai 200433, China}%
\affiliation{Shanghai Research Center for Theoretical Nuclear Physics, NSFC and Fudan University, Shanghai 200438, China}

\author{Yu-Gang Ma}
\affiliation{Key Laboratory of Nuclear Physics and Ion-beam Application (MOE), Institute of Modern Physics, Fudan University, Shanghai 200433, China}%
\affiliation{Shanghai Research Center for Theoretical Nuclear Physics, NSFC and Fudan University, Shanghai 200438, China}
\affiliation{School of Physics, East China Normal University, Shanghai 200241, China}

\maketitle

\section{The Coulomb potential with finite-size corrections}
\label{supp:coulomb-form}

For two point-like charged particles
with charges $Z_1$ and $Z_2$,
the Coulomb interaction is given by
\begin{equation}
    \label{eq:supp-coulomb}
    V_{\rm C}(r)=\frac{Z_1Z_2\alpha}{r}.
\end{equation}
However, real hadrons have finite charge radii, $r_d$, 
and finite charge distributions.
Therefore, the point-charge $1/r$ singularity is regularized and the potential is modified in the region $r\lesssim r_d$, while for $r\gg r_d$ the point-charge Coulomb potential  must be recovered. We therefore parametrize the finite-size-corrected potential as
\begin{equation}
    \label{eq:supp-coulomb-FC2}
    V(r)=\frac{Z_1Z_2\alpha}{r}F_{\rm C}(r),
\end{equation}
where the dimensionless function $F_{\rm C}(r)$ satisfies $F_{\rm C}(r)\to 1$ as $r\to\infty$. For the $p\Omega$ system, $Z_1Z_2=-1$. 

We describe the charge distribution of each hadron by a dipole form factor~\cite{Can:2015exa} in momentum space,
\begin{equation}
    \label{eq:supp-dipole-FF}
    F(k)=Q\frac{\Lambda^4}{(k^2+\Lambda^2)^2},
\end{equation}
where the parameter $\Lambda$
is related to the charge radius as shown below. 
The corresponding charge density is obtained by the Fourier transform of Eq.~\eqref{eq:supp-dipole-FF},
\begin{equation}
    \label{eq:supp-charge-density}
    \rho(r)=Q\frac{\Lambda^3}{8\pi}e^{-\Lambda r},
\end{equation}
which is normalized as $\int d^3r\,\rho(r)=Q$. The charge radius is defined by $\langle r_E^2\rangle=Q^{-1}\int d^3r\,r^2\rho(r)=12/\Lambda^2$, so that
\begin{equation}
    \label{eq:supp-charge-radius}
    \begin{aligned}
        r_d&=|\langle r_E^2\rangle|^{1/2},\\
        \Lambda&=\frac{2\sqrt{3}}{r_d}.
    \end{aligned}
\end{equation}
We use the adopted proton charge radius $r_d=0.841$~fm~\cite{Mohr:2024kco}, consistent with recent hydrogen spectroscopy determinations~\cite{Maisenbacher:2026nau,Bullis:2026ebw}, and for the $\Omega$ baryon the lattice result 
$r_d=0.571(18)$~fm~\cite{Can:2015exa}.

The interaction of two hadrons is obtained by convolving the point Coulomb potential with the two charge densities $\rho_1$ and $\rho_2$,
\begin{equation}
    \label{eq:supp-coulomb-conv}
    V(r)=\alpha\int d^3r_1\,\rho_1(r_1)\int d^3r_2\,\rho_2(r_2)\frac{1}{|\vec{r}-\vec{r}_1+\vec{r}_2|}.
\end{equation}
Using the momentum-space representation of the Coulomb kernel, the angular integration can be carried out analytically, and the result takes the form of Eq.~\eqref{eq:supp-coulomb-FC2}.

For equal charge radii, $\Lambda_1=\Lambda_2=\Lambda$, one obtains
\begin{equation}
    \label{eq:supp-FC-equal}
    F_{\rm C}(r)=1-e^{-\Lambda r}\left(1+\frac{11}{16}\Lambda r+\frac{3}{16}\Lambda^2r^2+\frac{1}{48}\Lambda^3r^3\right).
\end{equation}
For unequal charge radii, $\Lambda_1\neq \Lambda_2$, setting $a=\Lambda_1^2$ and $b=\Lambda_2^2$, one obtains
\begin{equation}
    \label{eq:supp-FC-unequal}
    \begin{aligned}
        F_{\rm C}(r)=\Lambda_1^4\Lambda_2^4\bigg[
        &-\frac{2(1-e^{-\Lambda_1 r})}{a(b-a)^3}+\frac{2(1-e^{-\Lambda_2 r})}{b(b-a)^3}\\
        &+\frac{2(1-e^{-\Lambda_1 r})-\Lambda_1 r e^{-\Lambda_1 r}}{2a^2(b-a)^2}\\
        &+\frac{2(1-e^{-\Lambda_2 r})-\Lambda_2 r e^{-\Lambda_2 r}}{2b^2(b-a)^2}\bigg].
    \end{aligned}
\end{equation}
Both expressions tend to unity as $r\to\infty$ and are smaller than unity at finite $r$, so that the finite-size correction slightly weakens the Coulomb attraction at short distances, while the point-charge Coulomb form is recovered at long distances. In this work, the proton and the $\Omega$ baryon have different charge radii, and we use the unequal-radius expression in Eq.~\eqref{eq:supp-FC-unequal}.

\section{Statistical analysis of the pole momentum}
\label{supp:bootstrap}

\begin{figure}[tb]
    \includegraphics[width=0.99\linewidth]{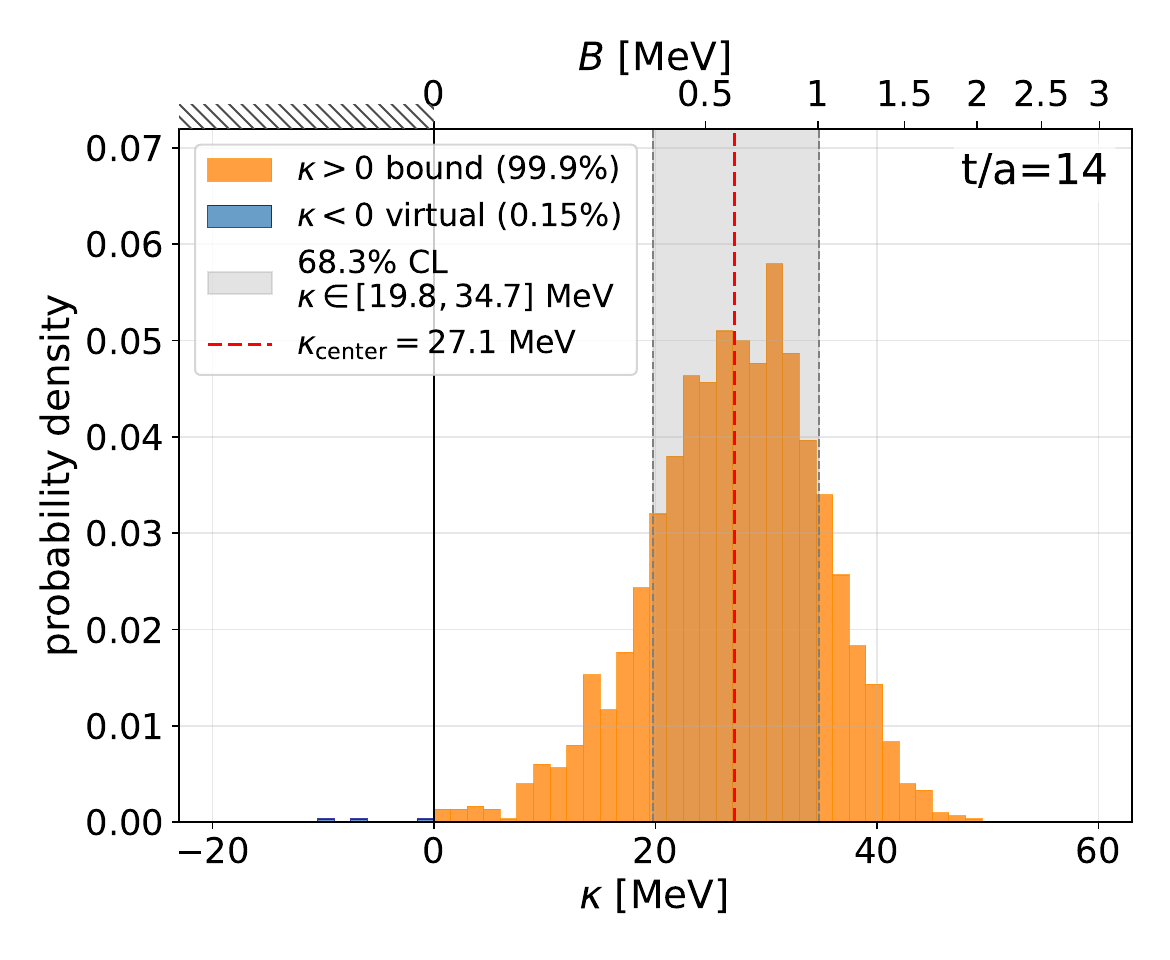}
    \caption{The pole momentum distribution from bootstrap resampling for the $^5S_2~n\Omega$ system at the reference time slice $t/a=14$. The distribution is approximately symmetric about its central value. The bound/virtual boundary ($\kappa = 0$) lies below the central 68.3\% ($1\sigma$) interval, so the $1\sigma$ interval lies entirely within the bound-state region ($\kappa>0$), indicating a bound $n\Omega$ system.}
    \label{fig:supp-pole-dis-nOmg}
\end{figure}

\begin{figure}[tb]
    \includegraphics[width=0.99\linewidth]{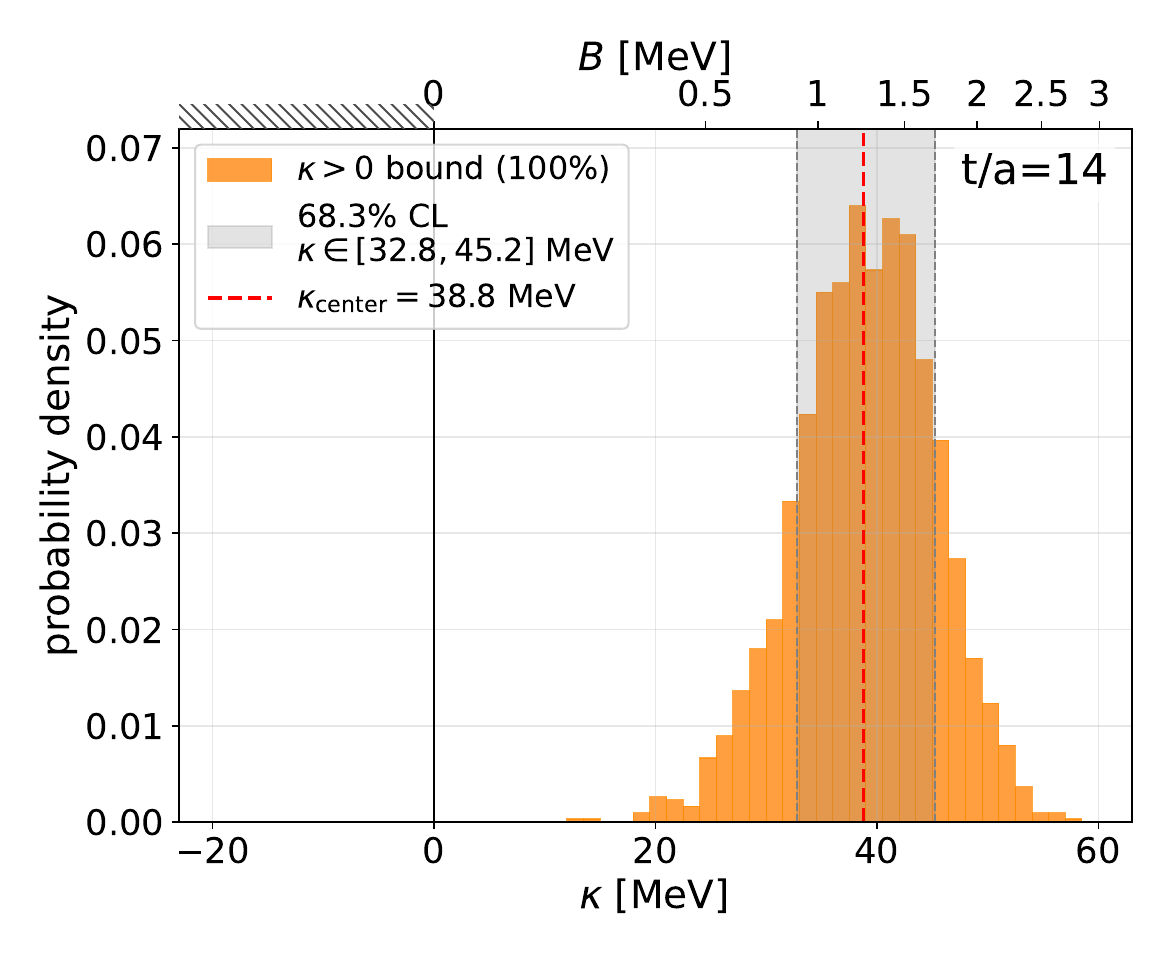}
    \caption{Same as Fig.~\ref{fig:supp-pole-dis-nOmg} but for the $^5S_2~p\Omega$ system at the reference time slice $t/a=14$. Due to the additional Coulomb attraction, all bootstrap samples 
    yield a bound-state pole for the $p\Omega$ system.
    }
    \label{fig:supp-pole-dis-pOmg}
\end{figure}

\begin{figure}[htbp]
    \includegraphics[width=0.99\linewidth]{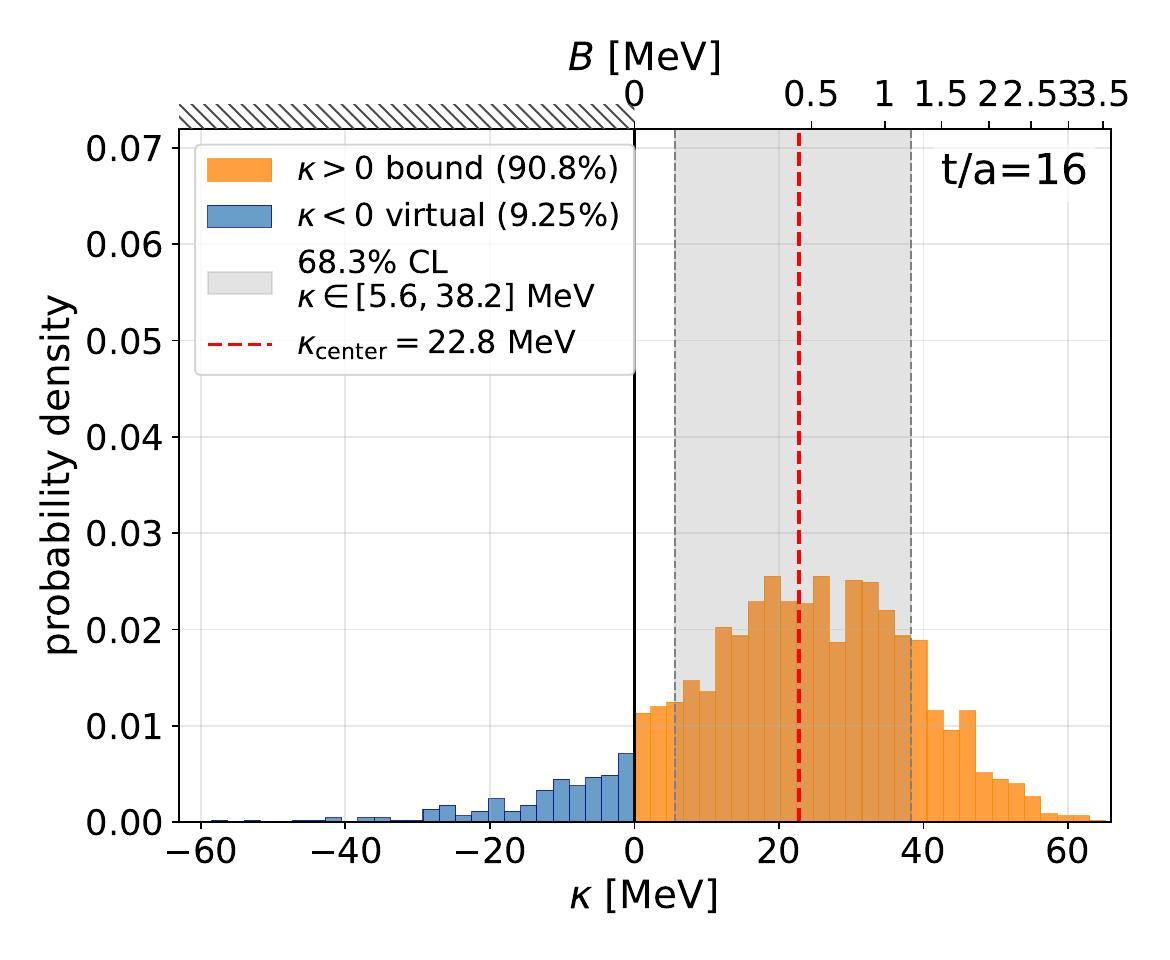}
    \caption{Same as Fig.~\ref{fig:supp-pole-dis-nOmg} 
    but at the time slice $t/a = 16$.
    The potential at this time slice has a relatively large statistical uncertainty, and about 9\% of bootstrap samples yield a virtual-state pole.}
    \label{fig:supp-pole-dis-nOmg-t16}
\end{figure}

Since the $^5S_2~N\Omega$ bound state is found to be close to the threshold in this work,
the binding energy $B$ is no longer a convenient quantity for a statistical analysis 
because statistical fluctuations may shift the near-threshold pole between a bound state and a virtual state, 
whereas $B$ is defined only for a bound state.
Consequently, estimating the statistical uncertainty of $B$ using only samples with a bound-state pole may introduce a bias.

We therefore characterize this near-threshold state through the pole momentum $\kappa$,
which is defined by the pole conditions in Eq.~\eqref{eq:fredholm} for $n\Omega$ and its Coulomb-modified counterpart for $p\Omega$, evaluated at $k=i\kappa$.
A single signed $\kappa$ covers both bound 
and virtual states:
A pole on the upper plane  (bound state) has $\kappa>0$, 
while a pole on the lower plane (virtual state) has $\kappa<0$.
This allows all statistical samples to be treated on an equal footing.

To estimate the statistical distribution of $\kappa$ and to check the jackknife error analysis used
in the main text, we generate $N_{\rm boot}=2000$ bootstrap samples from the 20 binned samples of the potential.
At the reference time slice $t/a=14$, the distributions of $\kappa$ are shown in Fig.~\ref{fig:supp-pole-dis-nOmg} for the $n\Omega$ system and in Fig.~\ref{fig:supp-pole-dis-pOmg} for $p\Omega$.
The central values and the $68.3\%$ intervals for $\kappa$ are
\begin{equation}
    \label{eq:supp-bootstrap-pole-pos}
    \begin{aligned}
        \kappa_{{n\Omega}} = 27.1(^{+7.6}_{-7.4 })~ \text{MeV},\quad
        \kappa_{{p\Omega}} = 38.8(^{+6.4}_{-6.0})~\text{MeV}.
    \end{aligned}
\end{equation}
The upper and lower half-widths are nearly equal for both channels, so each distribution is approximately symmetric about its central value. 
The bootstrap estimates of $\kappa$ are in good agreement 
with the jackknife results quoted in Eq.~(\ref{eq:pole-pos}).
We also confirm that
the scattering parameters $a_0$ and $r_{\rm eff}$ obtained from the same bootstrap samples agree with jackknife results.

\begin{figure*}
    \includegraphics[width=0.49\linewidth]{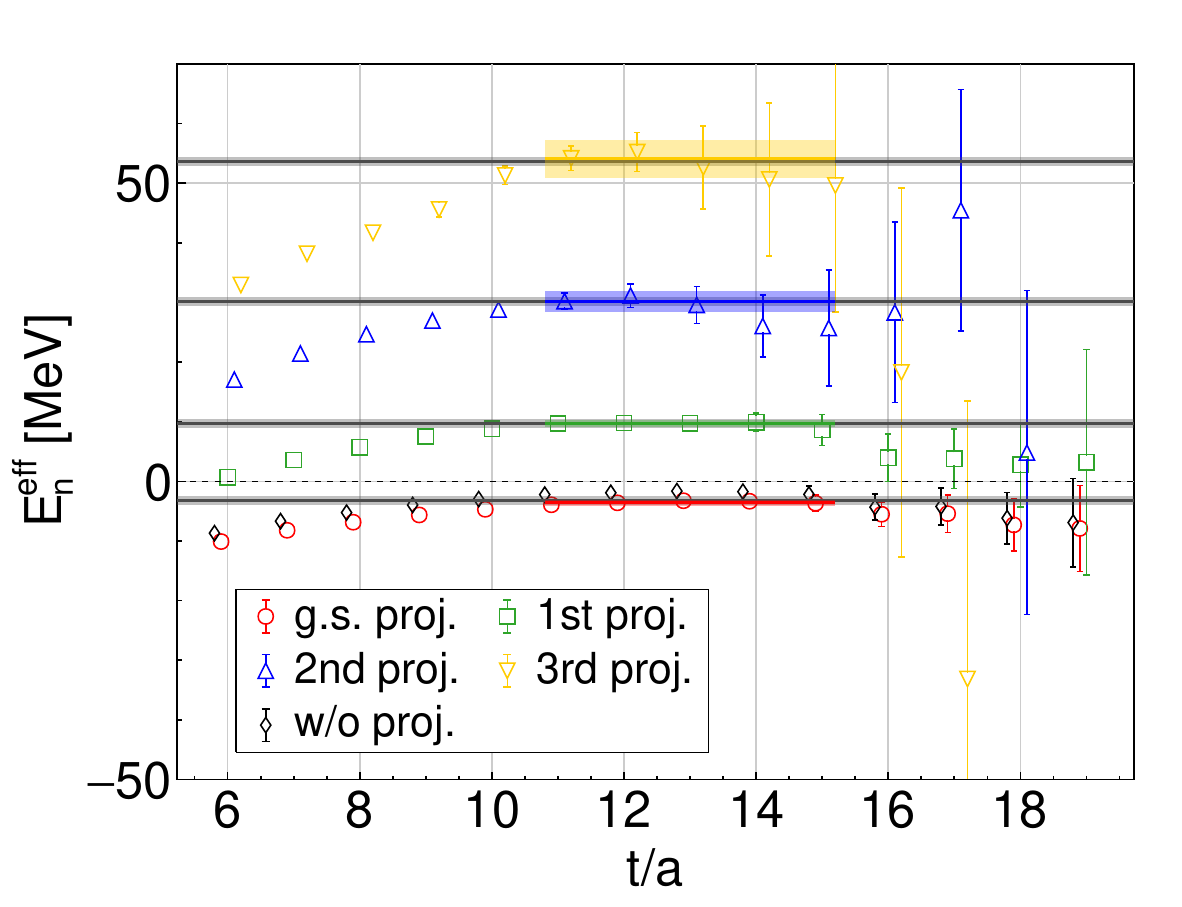}
    \includegraphics[width=0.49\linewidth]{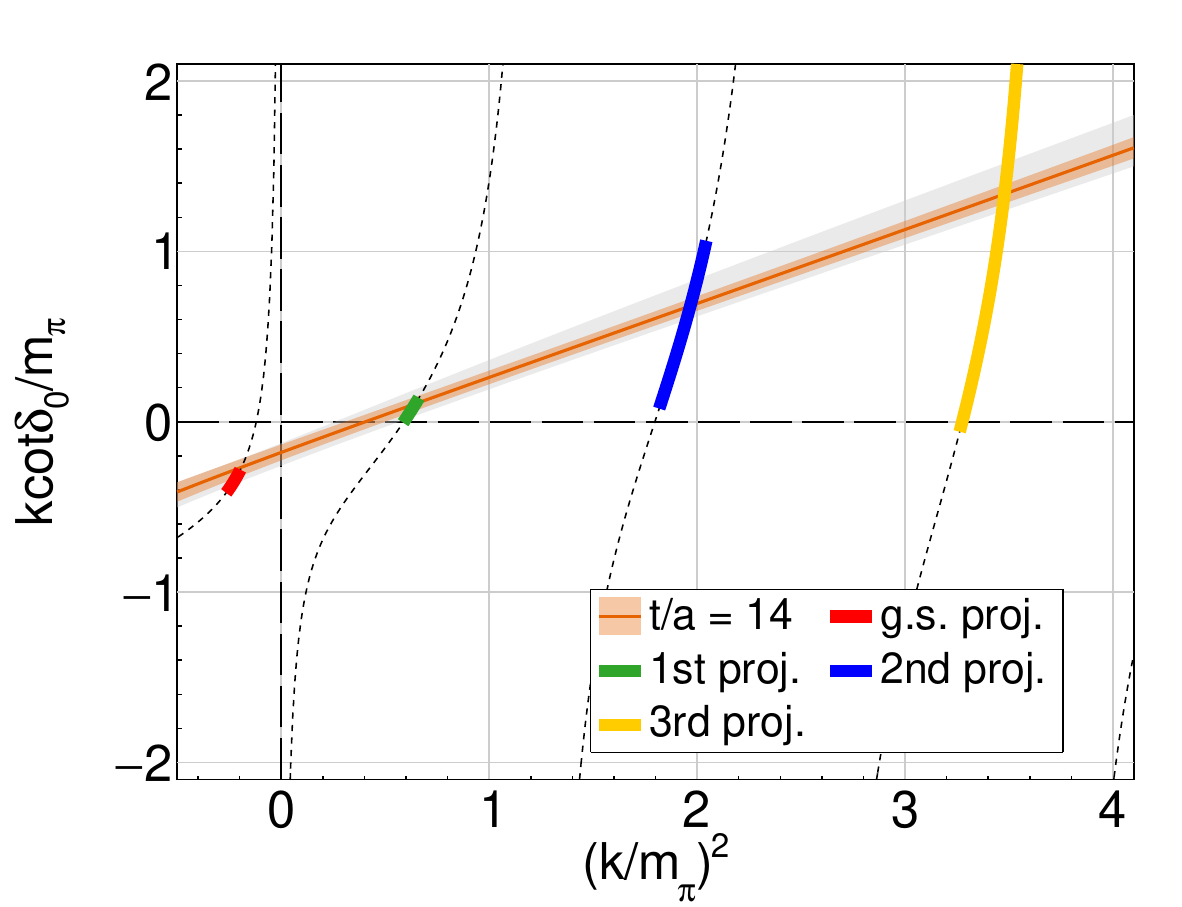}
    \caption{(Left) The four lowest eigenvalues $E_n'$ of the Hamiltonian $H'$, constructed from the LO potential extracted via the HAL QCD method at $t/a=14$ (gray bands), compared with the effective energies $E_n^{\rm eff}$ obtained by projecting the corresponding eigenfunctions $\phi_n(\vct{r})$ onto the $R$-correlator [Eq.~\eqref{eq:supp-proj}]. The black diamonds represent the effective energies extracted from the unprojected temporal correlator, $R_{\rm unproj}(t)=\sum_{\vct{r}}R(\vct{r},t)$. 
    (Right) The colored segments show $k\cot\delta_0$ inferred from the finite-volume effective energies using L\"uscher's relation, compared with $k\cot\delta_0$ (and its analytic continuation at $k^2<0$) calculated from the LO potential via the infinite-volume $T$-matrix (orange line with orange and gray bands for statistical and total uncertainties, respectively). 
    }
    \label{fig:supp-FiniteV}
\end{figure*}

Since $\kappa > 0$ corresponds to a bound state,
the fraction of bootstrap samples supporting a bound-state pole
in the $^5S_2~N\Omega$ system can be determined
from the distributions of $\kappa$.
For $n\Omega$,
at least $90\%$ of the samples support a bound state and the $1\sigma$ interval of $\kappa$ lies entirely in the region $\kappa>0$ at every time slice, $t/a=12$--$16$.
See Fig.~\ref{fig:supp-pole-dis-nOmg-t16} for $t/a=16$,
where the statistical fluctuations are largest and the fraction of bound-state samples is smallest.
For $p\Omega$, all samples yield a bound-state pole at every time slice. Within the present single-channel approximation, the resampling results favor a shallow bound state over a virtual state.

\section{Consistency check using finite-volume energy levels}
\label{app:FV-energy}

To check the consistency of the leading-order (LO) HAL QCD potential 
with the finite-volume energy spectrum, we compare energy levels predicted by the potential with those extracted from lattice temporal correlation functions.
The $R$-correlator [Eq.~\eqref{eq:R-corr}] can be expanded in terms of the 
eigenstates as~\cite{Ishii:2006ec,Aoki:2009ji}
\begin{equation}
    \label{eq:supp-R-corr-expand}
    R(\vct{r},t)=\sum_n A_n \psi_n(\vct{r})e^{-E_n t},
\end{equation}
where $\psi_n(\vct{r})$ and $E_n$ are the 
corresponding Nambu-Bethe-Salpeter (NBS) wave function and 
the eigenenergy measured relative to the $N\Omega$ threshold, respectively.
If the eigenfunctions $\psi_n(\vct{r})$ are known, the individual contributions to the $R$-correlator can, in principle, be isolated, allowing the energy spectrum of the two-baryon system to be extracted. 
In practice, the potential obtained by the HAL QCD method can be used to construct a discrete Hamiltonian $H'$ in the finite lattice volume. Its eigenfunctions
 provide an approximation to 
$\psi_n(\vct{r})$~\cite{Iritani:2018vfn,Lyu:2022tsd,Lyu:2025lnd,Lyu:2025ncq}.

We construct $H'$ using the LO potential [Eq.\eqref{eq:Vc}], with the Laplacian evaluated by finite differences between neighboring lattice sites. We then employ the Lanczos algorithm to calculate the lowest few eigenvalues ($E'_n$) and their corresponding eigenfunctions ($\phi_n(\vct{r})$). 
Projecting the $R$-correlator onto $\phi_n(\vct{r})$ yields
\begin{equation}
    \label{eq:supp-proj}
    \begin{aligned}
        R_n(t)&=\frac{1}{L^3}\sum_{\vct{r}} \phi_n^*(\vct{r})R(\vct{r},t)\\
              &\approx A_n e^{-E_n t}.
    \end{aligned}
\end{equation}
This projection approximately isolates the contribution from an individual state and corresponds to constructing an optimized two-baryon operator 
at the sink~\cite{Iritani:2018vfn, Lyu:2022tsd}.
(See also Refs.~\cite{Lyu:2025lnd,Lyu:2025ncq}, where $\phi_n(\vct{r})$ is used to 
construct the optimized operators at the source as well.)
The corresponding effective energy is then defined as
\begin{equation}
    \label{eq:supp-Eeff}
    E^{\rm eff}_n(t)=\frac{1}{a}\log\left(\frac{R_n(t)}{R_n(t+a)}\right).
\end{equation}
If $H'$ provides a good approximation to the underlying two-baryon Hamiltonian, the eigenvalues $E_n'$ should be consistent with the corresponding effective energies $E_n^{\rm eff}$~\cite{Lyu:2025lnd}.

For the $n\Omega$ system, we use the eigenfunctions of $H'$, constructed from the $t/a=14$ potential. The four lowest effective energy levels in the finite lattice volume obtained 
from $R_n(t)$ are
\begin{equation}
    \label{eq:supp-En}
    \begin{aligned}
        E^{\rm eff}_{\rm g.s.} &= -3.60\pm0.53~\text{MeV},\\
        E^{\rm eff}_{\rm 1st} &= 9.76\pm0.60~\text{MeV},\\
        E^{\rm eff}_{\rm 2nd} &= 30.16\pm1.77~\text{MeV},\\
        E^{\rm eff}_{\rm 3rd} &= 54.11\pm3.15~\text{MeV}.
    \end{aligned}
\end{equation}
As shown in Fig.~\ref{fig:supp-FiniteV} (Left),
these levels agree with the corresponding eigenvalues of $H'$ within uncertainties. 
This agreement provides a nontrivial consistency check of the LO potential for the low-energy spectrum, although it does not directly quantify all higher-order contributions.
Note that the finite-volume ground-state energy shift should not be identified with the infinite-volume binding energy in Eq.~\eqref{eq:binding-eng},
since the characteristic length $1/\kappa_{n\Omega}\simeq7.3$~fm is comparable to the spatial box size $L\simeq8.1$~fm.

Furthermore, to demonstrate the consistency in terms of the scattering observables, 
we calculate scattering phase shifts
from the finite-volume effective energies $E_n^{\rm eff}$
using L\"uscher's formula~\cite{Luscher:1990ux}.
As shown in Fig.~\ref{fig:supp-FiniteV} (Right), the resulting $k\cot\delta_0$ values (colored segments) agree with the $k\cot\delta_0$ curve extracted directly from the potential in infinite volume. 
This comparison further supports the LO description of the low-energy scattering observables.



\end{document}